\documentclass[fleqn,12pt,twoside]{article}
\usepackage{espcrc1,epsfig}

\usepackage{graphicx}
\usepackage[figuresright]{rotating}

\usepackage{bm}

\title{Using the $K^+p\to \pi^+K^+n$
reaction to determine the $\Theta^+$ quantum numbers}

\author{E. Oset,\address[MCSD]{Departmento de F\'isica Te\'orica and IFIC,
Centro Mixto Universidad de Valencia-CSIC,\\
       Institutos de Investigaci\'on de Paterna, Aptd. 22085, 46071
Valencia, Spain}%
	T.~Hyodo\address{Research Center for Nuclear Physics (RCNP),
Ibaraki, Osaka 567-0047, Japan}
    and 
    A.~Hosaka\addressmark  } 
\begin{document}

\date{\today}
\maketitle

\begin{abstract}
    We study the $K^+p\to \pi^+K^+n$ reaction with some kinematics
    suited to the production of the $\Theta^+$ resonance recently
    observed and show that the measurement of cross sections and polarization
    observables can shed light on the spin,isospin and parity of the $\Theta^+$
    state.
    
\end{abstract}

\section{ The $K^+p\to \pi^+K^+n$ reaction and the $\Theta^+$ quantum numbers}
A recent experiment by LEPS collaboration
at SPring-8/Osaka~\cite{Nakano:2003qx}
has found a clear signal for an $S=+1$ positive charge resonance
around 1540 MeV.
The finding, also confirmed by DIANA at
ITEP~\cite{Barmin:2003vv},
CLAS at Jefferson Lab.~\cite{Stepanyan:2003qr}
and SAPHIR at ELSA~\cite{Barth:2003es},
might correspond to the exotic state predicted by Diakonov {\em et~al.}
in Ref.~\cite{Diakonov:1997mm}.

The challenge now is 
to
determine the $\Theta^+$ quantum numbers through some
reaction.
We present one particularly suited reaction ~\cite{Hyodo:2003th} with the process
\begin{equation}
    K^+p\to \pi^+K^+n \ .
    \label{eq:reaction}
\end{equation}
which has some peculiar features since there are no resonances in the initial
state and by choosing small momenta of the $\pi^+$, we shall be also far
away from the $\Delta^+$ resonance and hence the $K^+n(\Theta^+)$
resonance signal can be more clearly seen.

\begin{figure}[tbp]
    \centering
    \includegraphics[width=9cm,clip]{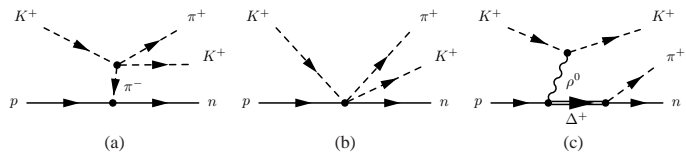}
    \caption{\label{fig:1}
    Feynman diagrams of the reaction $K^+p\to \pi^+K^+n$
    in the model of Ref.~\cite{Oset:1996ns}.}
\end{figure}

\begin{figure}[tbp]
    \centering
    \includegraphics[width=8cm,clip]{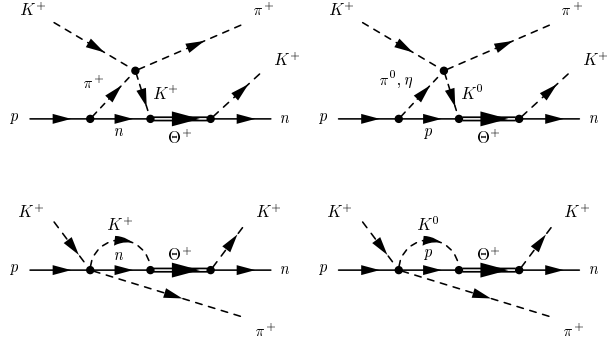}
    \caption{\label{fig:2}
    Feynman diagrams of the reaction $K^+p\to \pi^+K^+n$
    with the $\Theta^+$ resonance.}
\end{figure}

A successful model for the reaction Eq. (\ref{eq:reaction}) was
considered in
Ref.~\cite{Oset:1996ns}, consisting of the mechanisms depicted in
terms of Feynman diagrams in Fig.~\ref{fig:1}.
The term (a) (pion pole) and (b) (contact term) which are easily
obtained from the chiral Lagrangians involving
meson-meson~\cite{Gasser:1985gg}
and meson-baryon interaction~\cite{Meissner:1993ah}
are spin flip terms (proportional to $\bm{\sigma}$), while the $\rho$
exchange term (diagram (c)) contains both a spin flip and a non spin 
flip part. Having an amplitude proportional to $\bm{\sigma}$ is
important in the present context in order to have a test of the
parity of the resonance.
Hence we choose a situation, with the final pion momentum 
$\bm{p}_{\pi^+}$ small compared to 
the momentum of the initial kaon, such that the diagram (c), which 
contains the $\bm{S}\cdot \bm{p}_{\pi^+}$ operator can be safely
neglected.
The terms of Fig.~\ref{fig:1} (a) and (b) will provide the bulk for
this reaction.
If there is a resonant state for $K^+n$ then this 
will be seen in the final state interaction of this system.
This means that in addition to the diagrams (a) and (b) 
of Fig.~\ref{fig:1},
we shall have those in Fig.~\ref{fig:2}.
If the resonance is an $s$-wave $K^+n$ resonance then $J^P=1/2^-$.
If it is a $p$-wave resonance, we can have $J^P=1/2^+, 3/2^+$.

The restriction to have small pion momenta eliminates also other 
possible mechanisms involving crossed resonance exchange.

We write the couplings of the resonance to $K^+n$ as $g_{K^+n}$,
$\bar{g}_{K^+n}$ and $\tilde{g}_{K^+n}$ 
for $s$-wave and $p$-wave with $J^P=1/2^+,3/2^+$ respectively, and relate them
to the $\Theta^+$ width via

\begin{equation}
	g_{K^+n}^2
	=\frac{\pi M_R\Gamma}{Mq} \ , \quad
	\bar{g}_{K^+n}^2
	=\frac{\pi M_R\Gamma}{Mq^3} \ , \quad
	\tilde{g}_{K^+n}^2
	=\frac{3\pi M_R\Gamma}{Mq^3} \ .
    \label{eq:couplings}
\end{equation}

A straightforward evaluation of the meson pole and contact terms leads to the $K^+n\to \pi^+KN$
amplitudes
\begin{equation}
    -it_i
    =a_i\bm{\sigma}
    \cdot\bm{k}_{in}
    +b_i\bm{\sigma}
    \cdot\bm{q}^{\prime} \ ,
    \label{eq:t1amp}
\end{equation}
where $i=1,2$ stands for the final state $K^+n, K^0p$ respectively
and $k_{in}$ and $q^{\prime}$ are the initial and final $K^+$ momenta.

Now let us turn to the resonance diagrams of Fig.~\ref{fig:2}
containing a loop integral, which is initiated by the tree diagrams
of Figs.~\ref{fig:1} (a) and (b).
When taking into account $KN$ scattering through the $\Theta^+$
resonance, as depicted in Fig.~\ref{fig:2}, the $K^+p\to \pi^+K^+n$
amplitude is given by
\begin{equation}
    -i\tilde{t}=-it_1-i\tilde{t}_1-i\tilde{t}_2
    \label{eq:total}
\end{equation}
where $\tilde{t}_1$ and $\tilde{t}_2$ account for the scattering
terms with intermediate $K^+n$ and $K^0p$, respectively.
They are given by
\begin{eqnarray}
    -i\tilde{t}^{(s)}_i
    &=&c_i
    \bm{\sigma}\cdot\bm{k}_{in} \ , \nonumber\\
    -i\tilde{t}^{(p,1/2)}_i
    &=&d_i
    \bm{\sigma}\cdot\bm{q}^{\prime}  \ ,
    \label{eq:tilde} \\
    -i\tilde{t}^{(p,3/2)}_i
    &=&f_i\bm{\sigma}\cdot\bm{k}_{in}
    -g_i\bm{\sigma}\cdot\bm{q}^{\prime} \ , \nonumber
\end{eqnarray}
for $s$- and $p$-wave, and $i=1,2$ for $K^+n$ and $K^0p$
respectively.

We take an initial three momenta of $K^+$ in the Laboratory frame
$k_{in}(Lab)=850$ MeV$/c$ 
($\sqrt{s}=1722$ MeV), which allows us to
span $K^+n$ invariant masses up to $M_I=1580$ MeV, thus covering
the peak of the $\Theta^+$, and still is small enough to have negligible
$\pi^+$ momenta with respect to the one of the incoming $K^+$.

\begin{figure}[tbp]
    \centering
    \includegraphics[width=9cm,clip]{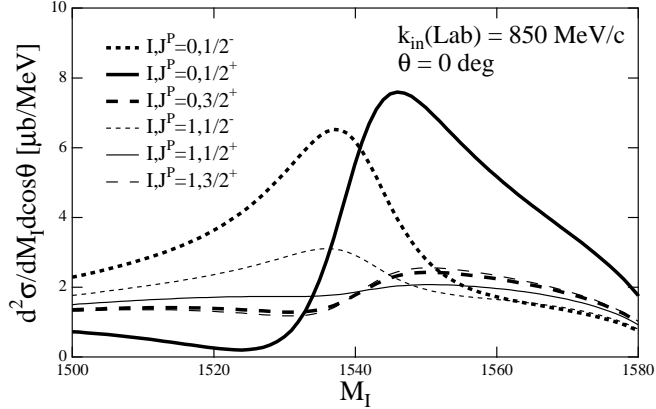}
    \caption{\label{fig:3}
    The double differential cross sections 
    $d^2\sigma/dM_Id\cos\theta$ with $\theta=0$ (forward direction)
    for $I=0,1$ and $J^P=1/2^-,1/2^+,3/2^+$.
    Below, detail of the lower part of the upper figure of the panel.}
\end{figure}

In Fig.~\ref{fig:3}, we show the invariant mass distribution
$d^2\sigma/dM_Id\cos\theta$ in the $K^+$
forward direction ($\theta=0)$.
Here we see that, independently of the quantum numbers of
$\Theta^+$, a resonance signal is always observed.
The signals for the resonance are quite clear for the case
of $I,J^P=0,1/2^+$ (these would be the quantum numbers predicted 
in Ref.~\cite{Diakonov:1997mm}) and  $I,J^P=0,1/2^-$, while in the other
cases the signal is weaker and the background more important,
particularly for the case of $I,J^P=0,1/2^+$. 

 The calculations are done with  a $\Theta ^+$ width of 20 MeV,
an 
experimental
upper bound. Should the width be smaller, the strength at the
peak of our calculation would be the same but the distributions would be
narrower.

Let us now see what can one learn with resorting to polarization
measurements. 
Eqs.~(\ref{eq:tilde})
account for the resonance contribution
to the process.
The interesting finding there is that if the $\Theta^+$ 
couples to $K^+n$ in $s$-wave (hence negative parity) 
the amplitude goes as $\bm{\sigma}\cdot \bm{k}_{in}$ 
while if it couples in $p$-wave it has a term
$\bm{\sigma}\cdot \bm{q}^{\prime}$.
Hence, a possible polarization test to determine which one of the
couplings the resonances chooses is to measure the cross section for
initial proton polarization $1/2$ in the direction $z$ $(\bm{k}_{in})$
and final neutron polarization $-1/2$  (the experiment
can be equally done with $K^0p$ in the final state, which makes the
nucleon detection easier).
In this spin flip amplitude $\langle -1/2 |t|+1/2\rangle$, the 
$\bm{\sigma}\cdot \bm{k}_{in}$ term vanishes.
With this test the resonance signal disappears for the $s$-wave
case, while the $\bm{\sigma}\cdot \bm{q}^{\prime}$ operator of the
$p$-wave case would have a finite matrix element 
proportional to $q^{\prime}\sin \theta$.

\begin{figure}[tbp]
    \centering
    \includegraphics[width=9cm,clip]{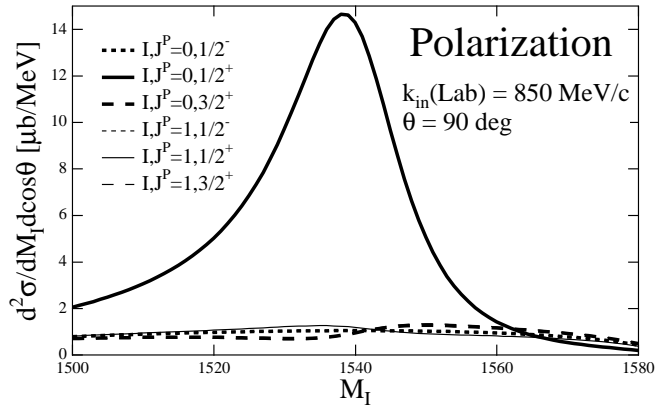}
    \caption{\label{fig:5}
    The double differential cross sections of polarized amplitude
    with $\theta=90$ for $I=0,1$ and $J^P=1/2^-,1/2^+,3/2^+$.
    Below, detail of the lower part of the upper figure of the panel.}
\end{figure}

In Fig.~\ref{fig:5} we show the results for the polarized cross section
measured at 90 degrees as a function of the invariant mass.
The two cases with s-wave do not show any resonant shape
since only the background contributes.
All the other cross sections are quite reduced to the point
that the only sizeable resonant peak comes from the $I,J^P=0,1/2^+$ case.
A clear experimental signal of the resonance in this observable
would unequivocally indicate the quantum numbers as  $I,J^P=0,1/2^+$.

\section*{Acknowledgments}
This work is supported by the Japan-Europe (Spain) Research
Cooperation Program of Japan Society for the Promotion of Science
(JSPS) and Spanish Council for Scientific Research (CSIC).
This work is also supported in part  by DGICYT
projects BFM2000-1326,
and the EU network EURIDICE contract
HPRN-CT-2002-00311.

%

\end{document}